\documentclass[a4paper,fleqn,usenatbib]{mnras}

\usepackage{newtxtext,newtxmath}

\usepackage[T1]{fontenc}
\usepackage{ae,aecompl}

\usepackage{graphicx}	
\usepackage{amsmath}	
\usepackage{amssymb}	

\newcommand{\unit}[1]{\,\mathrm{#1}}

\title[Understanding WASP 12b]{Understanding WASP 12b}

\author[A. Bailey and J. Goodman]{
Avery Bailey$^{1}$\thanks{E-mail: apbailey@astro.princeton.edu} and
Jeremy Goodman$^{1}$
\\
$^{1}$Department of Astrophysical Sciences, Princeton University, Princeton, NJ 08540, USA\\
}

\date{Accepted XXX. Received YYY; in original form ZZZ}

\pubyear{2017}

\begin{document}
\label{firstpage}
\pagerange{\pageref{firstpage}--\pageref{lastpage}}
\maketitle

\begin{abstract}
The orbital period of the hot Jupiter WASP-12b is apparently changing.  We study whether this reflects orbital decay
  due to tidal dissipation in the star, or apsidal precession of a slightly eccentric orbit.  In the latter case, a
  third body or other perturbation would be needed to sustain the eccentricity against tidal dissipation in the planet
  itself.  We have analyzed several such perturbative scenarios, but none is satisfactory.  Most likely therefore, the
  orbit really is decaying.  If this is due to a dynamical tide, then WASP-12 should be a subgiant without a convective
  core as \citet{Weinberg+2018} have suggested.  We have modeled the star with the \textsc{mesa} code.  While no model fits all
  of the observational constraints, including the luminosity implied by the GAIA DR2 distance, main-sequence models are
  less discrepant than subgiant ones.
\end{abstract}

\begin{keywords}
planet--star interactions -- stars: individual: WASP-12 -- planets and satellites: individual: WASP-12b
-- planets and satellites: dynamical evolution and stability
\end{keywords}



\section{Introduction}
Much circumstantial evidence indicates that tidal dissipation sculpts the orbits of short-period binary stars and
exoplanets.  First-principles tidal theories often have difficulty explaining the observations quantitatively, however.
For example, among low-mass main-sequence binaries, the period below which orbits circularize appears to increase with
system age up to periods $\sim 20\unit{d}$, whereas standard dissipation mechanisms become ineffective beyond
$\sim 10\unit{d}$ \citep{Zahn2013}.

Transiting exoplanets offer the prospect of testing tidal dissipation in real time.  Massive exoplanets with very short periods
are expected to exhibit orbital decay due to tidal dissipation in their host stars, whose rotation is usually
sub-synchronous, on timescales short compared to the star's main-sequence lifetime \citep{Levrard+2009}.  
(This should not occur for stellar binaries because of the much greater angular momentum in the orbit, only a small fraction of
which is needed to bring the stars into synchronous rotation.)  In favorable cases where the inspiral time is $\lesssim
10^7\unit{yr}$, transit timing with sub-minute accuracy may be expected to detect the period change after a decade or so.

Currently the most promising tentative detection has been made for WASP-12b, a planet with mass $m_{\mathrm b}\approx 1.5\ M_{\rm J}$ 
in a $1.0914\,\mathrm{d}$ orbit around a main-sequence F star \citep{Hebb+2009}.    Highly statistically
significant departures from a linear transit ephemeris have been measured by \citet{Maciejewski+2016} and
recently confirmed by \citet{Patra+2017}.   According to the latter authors, the measured rate of change of orbital
period is $\dot P = -29\pm 3\unit{ms\,yr^{-1}}$, and $P/\dot P=3.2\unit{Myr}$.

Three hypotheses for the orbital period change have been discussed.  One is orbital decay. A second is precession of the
periapse of a slightly eccentric orbit with a period $\sim 10\unit{yr}$ \citep{Maciejewski+2016}.  The required
eccentricity is on the order of $10^{-3}$, well below the limit $e<0.05$ set by \citet{Husnoo+2012}.  \citet{Patra+2017}
find that this explanation is disfavored by times of planetary occultation (secondary eclipse) as measured with {\it
  Spitzer}: an eccentric orbit would tend to displace the times of primary and secondary eclipses in opposite
directions, whereas the data seem to prefer an advance of both.
Furthermore, it seems unlikely that even such a small eccentricity could have survived tidal dissipation in the planet.
Nevertheless, \citet{Patra+2017} conclude that apsidal precession cannot yet be definitively ruled out on the basis of
the timing data.

The third possible explanation for $\dot P$ is acceleration
by a companion.  In fact WASP-12 is accompanied by a pair of M stars at projected separation $\approx 1\arcsec$,
\citep{Bechter+2014}.  Given that the estimated mass of this pair is $\approx 0.75\,M_\odot$ and the
distance to WASP-12 is $432.5\pm6.1\unit{pc}$ \citep{GaiaDR2}, the maximum line of sight acceleration is
$\approx 0.33\unit{m\,s^{-1}\,yr^{-1}}$, corresponding to $|\dot P|< 0.1\unit{ms\,yr^{-1}}$, far smaller than the
observed value.  More to the point---because there might be unseen massive planets closer in---\citet{Knutson+2014}
have used their radial-velocity data to place a limit $\lesssim 4\unit{m\,s^{-1}\,yr^{-1}}$ on this acceleration, and
this is still almost an order of magnitude too small to explain $\dot P$.

In the absence of a plausible fourth hypothesis, orbital decay would therefore seem to be the best explanation for the
observed departures from a linear ephemeris.  There are, however, reasons for doubt.  If the orbital decay timescale is
in fact only $\sim 3\unit{Myr}$, whereas the main-sequence lifetime of the host star is $\gtrsim 1\unit{Gyr}$ (see
\S\ref{sec:models}), we must be viewing the system at a special time.  On the other hand, WASP-12 is perhaps the best
current candidate for measurable orbital decay out of hundreds of hot Jupiters, so perhaps such a ``coincidence'' should
be less surprising.  A potential concern is the small measured rotation: $v\sin i<2.2\unit{km\,s^{-1}}$ \citep{Hebb+2009},
$v\sin i<5.1\unit{km\,s^{-1}}$ \citep{Fossati+2010b}, or $v\sin i=3.4\pm0.9\unit{km\,s^{-1}}$
\citep{Torres+Fischer+2012}.  If the planetary orbit has donated much of its original angular momentum to the star, one
might expect the star to have a larger $v\sin i$: the converse argument has been used by \citet{Penev+2016} to suggest
orbital decay in the HATS-18 system.  In \S\ref{sec:eq} and \S\ref{sec:spinup} however, we demonstrate that this expectation is 
incorrect and that tidal mechanisms are insufficient to bring WASP-12 to full synchronous rotation.  Instead, tidal mechanisms
should spin-up only a small core region of the star--the observational effect of which we explore.

The orbital decay explanation has been previously investigated by \citet{Weinberg+2018} who offer the novel suggestion
that WASP-12 is a subgiant star. 
But because this particular system holds a unique and valuable place within the context of tidal theories and planet-star interaction, 
we felt it necessary to investigate this system further.  We make a more thorough examination of stellar models before independently 
coming to similar interpretations as \citet{Weinberg+2018}.  Our analysis also benefits from the most recent luminosity estimates for WASP-12 
(see \S\ref{sec:lum}) and while the results are inconclusive, this new luminosity favors a higher mass main-sequence model.  
Bearing in mind this preference for a main-sequence model, we present a comprehensive investigation of alternative explanations
for the observed period change in \S\ref{sec:disc}.

\section{Stellar Models}\label{sec:models}
The tidal dissipation mechanisms discussed here are sensitive to the internal structure of the star, particularly the
existence and extent of convection zones.  Therefore, we begin by selecting a fiducial model for the WASP-12 host.

\subsection{Properties of the WASP-12 star}\label{subsec:properties}
\begin{table*}
  \centering
  \caption{Observed and adopted properties of WASP-12}
  \begin{tabular}{cccl}
    \hline
    $T_{\rm eff}$ & [Fe/H] & $\bar\rho$ & Reference \\
    K & dex & $\bar\rho_\odot$  & \\
    \hline
    $6300^{+200}_{-100}$ & $0.3^{0.05}_{0.15}$  & $0.35\pm0.03$ & \citet{Hebb+2009}\\[5pt]
    $6250\pm100$       & $0.32\pm0.12$     & ---                          & \citet{Fossati+2010b}\\
    $6118\pm 64$        & $0.07\pm0.07$     & ---                          & \citet{Torres+Fischer+2012}\\
    ---                          & ---                        & $0.325\pm0.016$   & \citet{Southworth2012}\\
    $6313\pm52$         & $0.21\pm0.04$     & ---                          & \citet{Mortier+2013}\\
    ---                          & ---                       &  $0.315\pm0.007$   & \citet{Maciejewski+2013}\\[10pt]
    $6241\pm36$         & $0.198\pm0.032$ & $0.3181\pm 0.0063$ & Adopted\\ 
    \hline
  \end{tabular}
  \label{tab:properties}
\end{table*}

Table~\ref{tab:properties} summarizes properties of WASP-12 as independently determined by the studies cited.
The effective temperature and metallicity are in principle directly determinable by comparison of spectra with
atmospheric models, while the mean stellar density $\bar\rho\equiv 3M_*/4\pi R_*^3$ follows from
the orbital period and the fractional width $R_*/a$ of the planetary transit.  We have chosen not to use
spectroscopic determinations of surface gravity, as some of these studies regard $\log g$ as
problematic unless constrained by the mean density.
The adopted values on the last line were obtained as a straightforward weighted average of the values shown:
\begin{equation}
  \label{eq:average}
  \hat X = \hat\sigma^{-2}\sum_k \frac{X_k}{\sigma_k^2},\qquad\hat\sigma^{-2}=\sum_k\frac{1}{\sigma_k^2}\,
\end{equation}
the original errors $\{\sigma_k\}$ being symmetrized where necessary, e.g. $6300^{+200}_{-100}\to6300\pm150$.
The adopted error $\hat\sigma$ is optimistic, especially for $T_{\rm eff}$ and [Fe/H], since the original
errors are probably dominated by systematics of the atmospheric models.

\subsection{Interior models}

Models for WASP-12 were constructed using the 10108 release of the publicly available 1-D stellar evolution code \textsc{mesa}
\citep{mesa1,mesa2,mesa3,mesa4}.  This version of \textsc{mesa} includes an improved prescription for determining radiative-convective boundaries,
the locations of which can sensitively alter the strength of tidal effects.
We arrived at a fiducial model for WASP-12 after searching the parameter space over mass $M_\ast$
and initial metallicity $Z_{\rm init}$ with bounds $1.15 \,M_\odot < M_\ast < 1.4 \,M_\odot$ and $0.01 < Z_{\rm init} < 0.033$.  
An unweighted sum of $\chi^2$ statistics of the adopted properties listed in Table~\ref{tab:properties} was chosen as the goodness-of-fit
statistic to be minimized over the searched parameter space.  
Though we conducted searches including all three observables in the goodness-of-fit statistic $\chi^2\left(\bar{\rho},T_{\rm eff},[\text{Fe/H}]\right)$, here
we focus on the results of searches for the statistic  $\chi^2\left(\bar{\rho},T_{\rm eff}\right)$ which omit [Fe/H].
Both statistics give similar results, but the latter lends itself to analyzing the observables in an individual sense instead of a combined one.
For our calculation of the model 
$[\text{Fe/H}]\equiv\log_{10}(Z_{\rm surf}/X_{\rm surf}) - \log_{10}(Z_\odot/X_\odot)$, we adopted the \citet{asp09} value, $Z_\odot/X_\odot = 0.0181$.  
For each combination of $M_\ast$ and $Z_{\rm init}$, a stellar model was evolved from pre-main sequence with our goodness-of-fit 
statistic evaluated at each timestep until the statistic moved far enough from a local minimum to trigger the stopping conditions for that evolutionary run.  

\begin{figure}
\centering
\includegraphics[width=\linewidth]{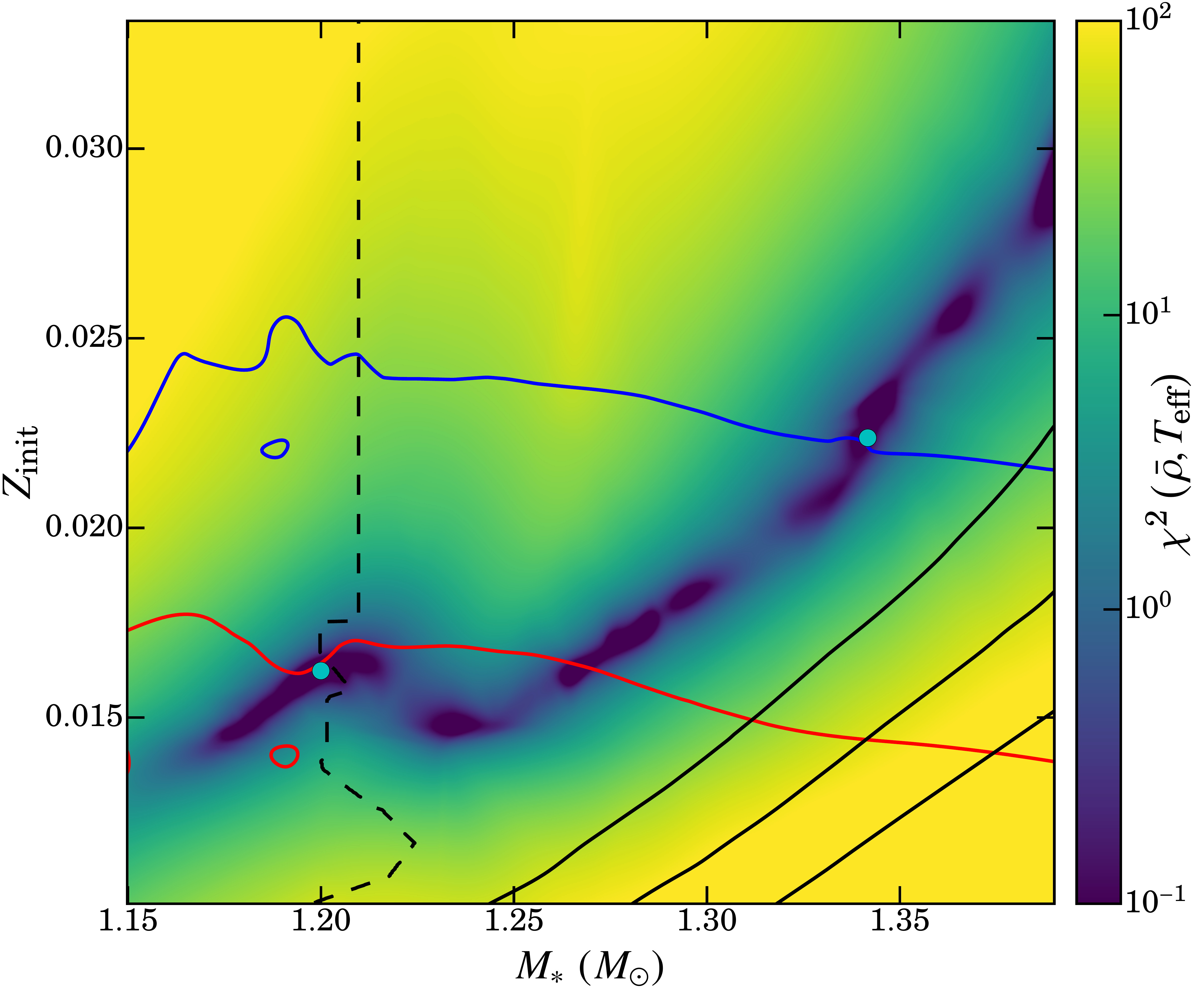}
\caption{The tested parameter space over initial mass and metallicity colored according to the goodness-of-fit statistic 
under a cubic interpolation.  Lines are plotted for the adopted [Fe/H]=0.198 (blue) and the solar value (red). Solid black contours for luminosity
are placed at the level: $\log_{10}(L_\ast/L_\odot)=0.65$ with $\pm 10\%$ error bars.
Models with masses
lower than the black dashed line have a radiative core whereas higher masses have a small convective core.  These models assume
mixing length parameter $\alpha=1.9$.  The location of models A \& B are marked as cyan points}
\label{fig:chisq1.9}
\end{figure}

\begin{figure}
\centering
\includegraphics[width=\linewidth]{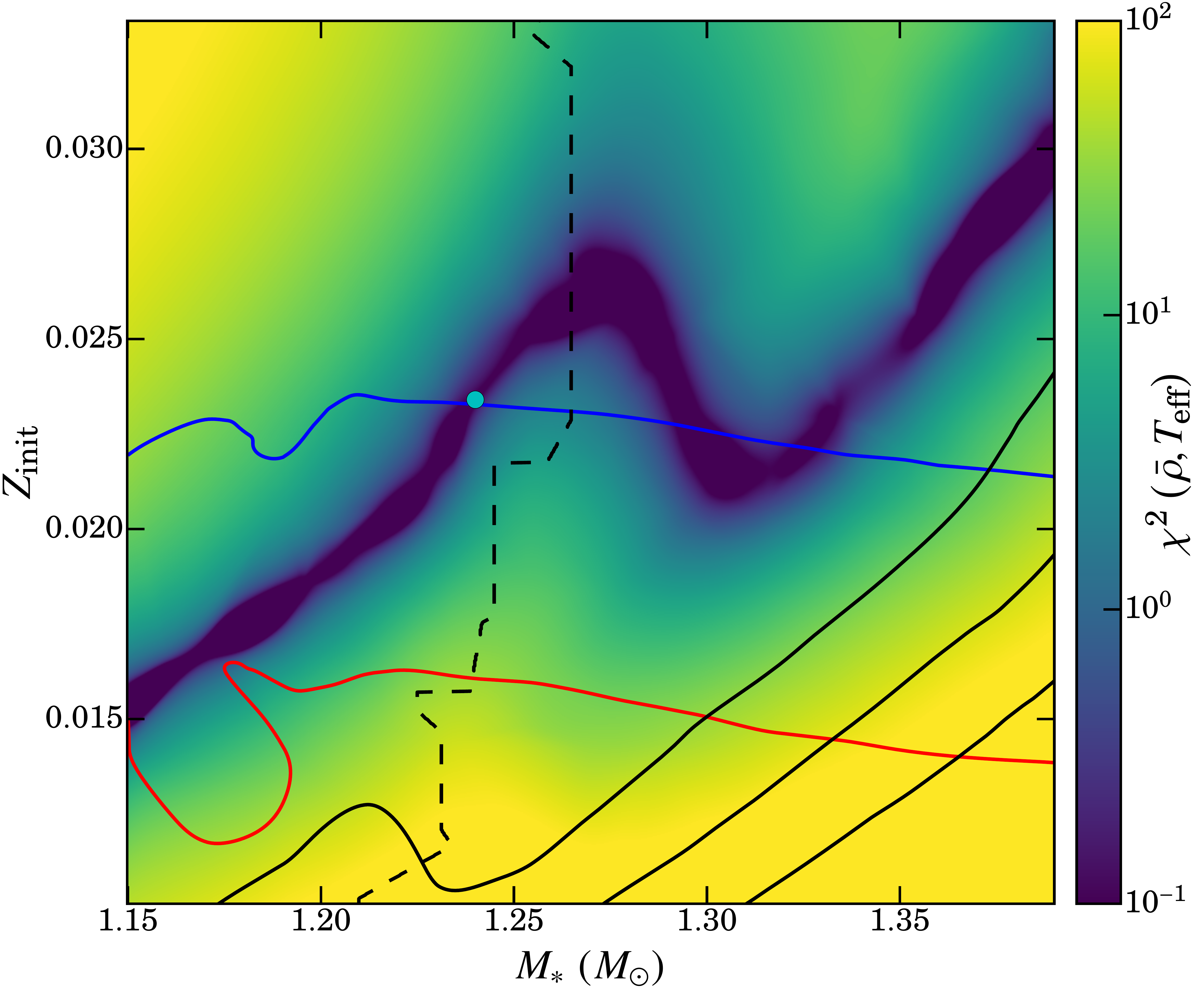}
\caption{The tested parameter space for models with mixing length parameter $\alpha=2.3$.  Plotted according to the caption in Fig. \ref{fig:chisq1.9}\label{fig:chisq2.3} with the location of model C marked as a cyan point}
\end{figure}

For all parameter space searches we adopted the physics of \citet{Choi2016}, but tested two mixing length parameter
values of $\alpha =1.9$ and $\alpha=2.3$.  The results of a series of evolutionary runs according to both several grid
and simplex searches for $\chi^2\left(\bar{\rho},T_{\rm eff}\right)$ are displayed in Fig. \ref{fig:chisq1.9} for
$\alpha=1.9$ and Fig. \ref{fig:chisq2.3} for $\alpha=2.3$.  We highlight three models in particular between
Figs. \ref{fig:chisq1.9} \& \ref{fig:chisq2.3} and provide additional details in Table \ref{tab:models}.  The first
model, model A, is representative of a class of models that are on the main-sequence, plus are able to adequately fit
the observed $\bar{\rho}$, $T_{\rm eff}$, and [Fe/H].  Most significantly, models of this type have a small convective
core.  If WASP-12 had no convective core, gravity waves excited at the outer convective--radiative boundary might
deposit their angular momentum by breaking non-linearly at the inner turning point where the Brunt--V\"{a}is\"{a}l\"{a}
$N$ equals the tidal frequency $\omega$ \citep{Goodman+Dickson1998,Terquem+1998,Barker+Ogilvie2010,Weinberg+2018}.  The
convective core in these models removes the possibility of this dissipation mechanism by moving the turning point
outward to a region where the gravity waves lack the amplitude to break.  As we show in $\S\ref{sec:eq}$ and
$\S\ref{sec:damp}$, models with the structure of model A are unable to explain the observed tidal decay of WASP-12b.

This motivated the search for an additional class of models that lack a convective core.
At lower masses, the convective core of models that still fit the observed $\bar{\rho}$ and $T_{\rm eff}$ shrinks.  
This continues until, as is displayed in Figs. \ref{fig:chisq1.9} \& \ref{fig:chisq2.3}, 
the convective core disappears entirely around a mass of $\approx1.2 \,M_\odot$ where the well-fitting models become subgiants.
We focus on two representative subgiant models that we refer to as model B and C, differentiated by having $\alpha =1.9$ and $\alpha=2.3$ respectively. 
Unfortunately, each of these models suffers from inconsistencies: model B having too low surface metallicity to match 
observations and model C having an unrealistically high $\alpha$.  

While our fiducial value for the error on [Fe/H] is undoubtably optimistic, past analyses have all concurred that WASP-12 
has supersolar surface abundance, whereas our model B is near solar or slightly subsolar. 
As the surface metallicity is dependent upon the prescription for elemental diffusion and stellar rotation, we also tested an alternative
prescription more tailored to WASP-12 than \citet{Choi2016}, assuming a rotation rate of $\approx 10\unit{km\,s^{-1}}$.  
This rotation rate was informed by measurements of the Rossiter--McLaughlin effect in this system that indicate a strong spin--orbit misalignment 
of $59^{+15\circ}_{-20}$ and that $v\sin i_\ast = 1.6^{+0.8}_{-0.4}$ km s$^{-1}$ \citep{2012ApJ...757...18A}.
Corresponding changes to the surface metallicity were minimal and it seems unlikely that any rotation or diffusion mechanism would be able
to significantly enhance the surface metallicity above $Z_{\rm init}$ to reproduce the observed [Fe/H].

Though model C fits well for all three observables, it assumes $\alpha=2.3$, in what is probably an unrealistically high choice of $\alpha$.
As Procyon is a spectral neighbor to WASP-12 and is particularly well-constrained, it provides reasonable calibration for $\alpha$.
Assuming a mass of $1.478\pm 0.012 M_\odot$ \citep{Bond2015}, setting initial [Fe/H] equal to the observed 
$\text{[Fe/H]}=-0.05\pm0.03$ \citep{Prieto2002}, models with $\alpha\geq2.2$ can not simultaneously reproduce the 
observed $T_{\rm eff}=6516 \pm 87$ K \citep{Auf2005}, luminosity $\log_{10}(L/L_\odot) =0.84 \pm 0.018$ \citep{Jerzy2000}, 
and $\bar{\rho}=0.1725 \pm 0.0007\,\rho_\odot$ \citep{Bedding2010}.
Instead, the best-fitting models occur in the range $1.8\leq \alpha \leq 2.1$.  Solar calibrations corroborate this, 
suggesting $\alpha=1.93$ \citep{2012ApJ...746...16V}, $\alpha = 1.82$ \citep{Choi2016}, etc.

While we certainly have not exhausted the parameter space of possible subgiant models, and there are enough tunable parameters
in stellar modeling that it may be possible to construct a subgiant model that fits the observables, with standard assumptions it is 
difficult to do so.  On the other hand, main-sequence models that fit the observables (except the luminosity---see
below) are generic and easy to find, which is likely why 
previous studies estimate the mass of WASP-12 to be near $1.4 M_\odot$ \citep{Collins+2017,Southworth2012}.  
Though the convective core in these models inhibits tidal decay via breaking of gravity waves, the presence of a convective 
envelope allows for damping of the equilibrium tide ($\S$\ref{sec:eq}) and 
dynamical tide ($\S$\ref{sec:damp}) by turbulent viscosity \citep{1977A&A....57..383Z}. 
In these later sections however, we adopt
model A as a fiducial model to show that both the equilibrium tide and dynamical tide in main-sequence models are unable to explain WASP-12b's decay.  

\begin{table*}
  \centering
  \caption{Fiducial WASP-12 Models}
  \begin{tabular}{ccccccccccc}
    \hline
    Name & $\alpha$ & $Z_{\rm init}$ & $M_\ast$ & Age & $\log_{10}(L_\ast/L_\odot)$ &$T_{\rm eff} $ 
    & [Fe/H] & $\bar{\rho}$ & $\chi^2\left(\bar{\rho},T_{\rm eff}\right)$ & $\chi^2\left(\text{[Fe/H]}\right)$\\
     & & & $M_\odot$ & Gyr & & K & dex & $\bar{\rho}_\odot$ \\
    \hline
A & 1.9 & 0.0223 & 1.34 & 2.72 & 0.55 & 6250 & 0.20 & 0.3182 & 0.064 & 0.031\\
B & 1.9 & 0.0162 & 1.20 & 4.24 & 0.52 & 6242 & -0.03 & 0.3185 & 0.006 & 49.06\\
C & 2.3 & 0.0234 & 1.24 & 4.51 & 0.53 & 6245 & 0.20 &  0.3181 & 0.017 & 0.006\\
    \hline
  \end{tabular}
  \label{tab:models}
\end{table*}

\subsection{WASP-12 luminosity}\label{sec:lum}
In addition to using $\bar{\rho}$ and $T_{\rm eff}$ to constrain stellar models, we made some investigation of whether our models A,B,C 
could be constrained via luminosity measurements.
The most recent parallax measurements made by the GAIA mission \citep{Gaia,GaiaDR2} place WASP-12 at a distance of $432.5\pm 6.1$ pc,
significantly further than past determinations.  Adapting the work of \citet{Stassun+2017}, who estimate the extinction to WASP-12 at $A_V=0.29$ mag., 
to this updated distance measurement gives a luminosity of $\log_{10}\left(L_\ast/L_\odot\right)= 0.65$.  
Taking $L_\ast$ together with our selected values of $T_{\rm eff}$ and $\bar{\rho}$ uniquely determines the mass at $M_\ast \approx 1.9 \,M_\odot$.  
Such a high mass would seem to favor the higher mass main-sequence models for WASP-12 but
evolutionary runs at $1.9 \,M_\odot$ fail to simultaneously fit these three observables. Figs. \ref{fig:chisq1.9} \& \ref{fig:chisq2.3} similarly suggest that any 
model significantly larger than $1.4 \,M_\odot$ and having the requisite $T_{\rm eff},\bar{\rho}$ would require an unrealistically high metallicity.  
For example, even the best-fitting highest metallicity $1.9\, M_\odot$ model tested ([Fe/H]$\approx 0.4$) had a value $\chi^2\left(\bar{\rho},T_{\rm eff},L_\ast\right) > 100$.
The incompatibility of these three observables is also visible in Figs. \ref{fig:chisq1.9} \& \ref{fig:chisq2.3} as lines of constant 
$L_\ast$ lie parallel to the track of constant $\bar{\rho},T_{\rm eff}$.  This tension between measured observables is alleviated as one goes to either higher
$T_{\rm eff}$ or lower luminosity.

While GAIA DR2 lists a very precise parallax for WASP-12, $2.3122\pm0.0325\,\mathrm{m.a.s.}$, the extinction ($A_G$) is
not reported.  Without correction for extinction, the reported luminosity is $3.435\pm 0.075\,\mathrm{L_\odot}$, which
is entirely compatible with the models in Table~\ref{tab:models}.  One might therefore worry that \cite{Stassun+2017}
have overestimated the extinction or the flux---the former perhaps because some of the photometric data they used were
published before it was recognized that the star has two M-dwarf companions within $1''$ \citep{2013MNRAS.428..182B}.
But for comparison, the dust map of \cite{Green+2018}\footnote{which can be queried at {\tt argonaut.skymaps.info}}
 predicts $E(B-V)=0.07^{0.02}_{0.03}\,\mathrm{mag}$ at 440~pc in
the direction of WASP-12, which would correspond to $A_V\approx 0.21\pm0.09\,\mathrm{mag}$ for a normal extinction curve.
Querying the Gaia~DR2 catalog for stars within one degree of WASP-12, parallaxes $\ge 2.28\,\mathrm{m.a.s.}$, and
$T_{\rm eff}>5500\,\mathrm{K}$ yields 101 results, of which 78 have $A_G$ values listed.  There is no clear trend with
distance, but the median $A_G$ for the more distant half of this sample is $0.23\,\mathrm{mag}$.
These are slightly lower than the \citet{Stassun+2017}  estimate, but consistent within the uncertainties.  So it seems
that WASP-12 is at least $\sim 10\mbox{\,-\,}30\%$ more luminous than any of the models in Table~\ref{tab:models}.

Considering these three independently determined observables ($\bar\rho,\,T_{\rm eff},\,L$) are incompatible with one
another, we also ran a chi-square model search including the luminosity, with $10\%$ errors on $L_\ast$.  As one can
infer from inspection of the luminosity contours in Figs. \ref{fig:chisq1.9} \& \ref{fig:chisq2.3}, including $L_\ast$
in the search moves the track of well-fitting models towards lower metallicity such that it lies between the low
$\chi^2(\bar{\rho},T_{\rm eff})$ track and the $\log_{10}(L_\ast/L_\odot) = 0.65$ contour.  As the three observables are
incompatible, $\chi^2(\bar{\rho},T_{\rm eff},L_\ast)$ becomes significantly more nonzero, with minimum $\chi^2\approx$
3--5, depending on $\alpha$ and assuming $10\%$ error bars on the luminosity.  Ultimately, it seems that because lines
of constant $\bar{\rho}, T_{\rm eff}$ lie nearly parallel to lines of constant $L_\ast$ in the $Z_{\rm init}-M_\ast$
plane, that luminosity measurements offer little guidance in choosing between models of type A, B, or C (at least in the
mass range $1.15\,M_\odot < M_\ast <1.4\,M_\odot$).

\section{Equilibrium Tide}\label{sec:eq}
The adiabatic equilibrium tide describes the hydrostatic tidal response of the host star to a perturbing body in the absence of dissipation.
In this hydrostatic limit where the tidal frequency $\omega$ goes to zero, the functional relationship between density $\rho$, pressure $p$, and 
potential $\Phi$ is preserved.  As a result, density and pressure are constant along equipotentials and have the same value on a 
given equipotential as they would on the same equipotential absent the tide.
If one neglects composition gradients, entropy $S$ is a function of $\rho$ and $P$ only and would consequently follow the equipotentials.  
In regions with a nonzero entropy gradient, i.e. radiative zones, the adiabatic condition would require fluid elements also stay tied to 
equipotentials ($\xi_r^{\rm eq} \propto \Phi_1$, the subscript $1$ indicating an Eulerian perturbation).
The equilibrium tide is also incompressible, $\pmb{\nabla} \cdot \pmb{\xi}^{\rm eq} = 0 $. In stably stratified regions the radial 
fluid displacement is explicitly described by the equation,
\begin{equation}\label{eq:eq}
\xi_r^{\rm eq} = \frac{\Phi_1}{d\Phi/dr}.
\end{equation}
With the addition of composition gradients, though the above entropy argument no longer holds,  a similar result can be derived.  
Namely, the fluid displacements are still described by eq. \eqref{eq:eq}
and are still incompressible where the square Brunt-V\"ais\"al\"a frequency $N^2  \neq 0 $ .
In convective regions where the entropy gradient vanishes, and $N^2 =0$, fluid displacements may not necessarily 
follow the above eq. \eqref{eq:eq} but one defines the equilibrium tide such that eq. \eqref{eq:eq} is satisfied.

In convective regions of a star, turbulent viscous forces facilitate the cascading of bulk kinetic energy to smaller scales where it is dissipated.
The form of this dissipative system and it's action on the equilibrium tide in stars with a convective region was developed by \citet{1966AnAp...29..489Z}
and dissipates the energy on a timescale \citep{2012A&A...544A.132R},
\begin{equation}\label{eq:tdiss}
\frac{1}{t_{\rm diss}} = 4\pi\frac{6264}{35}\frac{R_\ast}{M_\ast}\int_{R_{+}/R_\ast}^1 \rho\nu_t x^8 dx 
\end{equation}
where $x\equiv r/R_\ast$ is the fractional radius, $R_+$ is the radius of the outermost radiative-convective boundary, and $\nu_t$ is the convective viscosity.  
Here we have restricted the limits of integration to extend only over the convective envelope as the equivalent contribution 
due to the convective core is heavily suppressed by the $x^8$ dependence within the integrand.
Eq. \eqref{eq:tdiss} operates under the assumption of a thin convective envelope.  Stated more precisely:
\begin{enumerate}
\item the mass of the convective region is negligible ($M_\ast \approx M_{+}$),
\item the self-interacting perturbation to the potential caused by equilibrium tide displacements $\Phi_{1,\ast}$ is small 
compared to the perturbation to the potential caused by the star's external companion $\Phi_{1,\mathrm{b}}$  ($\Phi_{1,\ast}  + \Phi_{1,\mathrm{b}} \approx \Phi_{1,\mathrm{b}}$),
\item the stellar invariant $U$ is small compared to unity ($U\equiv d\ln M/ d\ln R \ll 1$).
\end{enumerate}
Our fiducial WASP-12 models satisfy the above criteria for a thin convective envelope with the convective envelope 
containing $< 0.2$ per cent of the mass of the star, $\Phi_{1,\ast} < 0.03 \Phi_{1,\mathrm{b}}$, 
and the stellar invariant $U  < 0.025$.

In calculating the dissipation rate associated with the equilibrium tide, we assume a viscosity of the form,
\begin{equation}\label{eq:visc}
\nu_t = \frac{l_c v_c}{\sqrt{1 + \left(\tau_c/P_{\rm tide}\right)^2}}
\end{equation}
where $l_c$ is the mixing length, $v_c$ is the r.m.s. vertical convective velocity, $\tau_c \equiv 2l_c/v_c$ is twice the local 
convective turnover time and $P_{\rm tide}=2\pi/\omega$ is the tidal period.  This form of the viscosity is a heuristic that reproduces 
the turbulent viscosity formalism of \citet{1966AnAp...29..489Z} in the limit that $P_{\rm tide} \gg \tau_c$ and $P_{\rm tide} \ll \tau_c$.  
The suppression of viscosity that arises in Zahn's formalism for $P_{\rm tide} \ll \tau_c$ comes from the fact that for large tidal frequencies, 
eddies are unable to travel a full mixing length.  One then supposes that the mean free path of such an eddy should be 
replaced by the distance an eddy travels in something like half a tidal period resulting in a suppression by a factor of $P_{\rm tide}/\tau_c$.  
Others such as \citet{1977Icar...30..301G} have argued that eddies with turnover times greater than the tidal period do not 
`exchange momentum with the mean flow on this time scale [the tidal period]' and therefore are not necessary in evaluating the diffusivity.  
The result is a suppression of the viscosity that is quadratic in $P_{\rm tide}/\tau_c$ rather than linear.  
This uncertainty in the form of the viscosity remains an outstanding problem in tidal theory but here we adopt Zahn's formalism partially 
because simulations done by \citet{2007ApJ...655.1166P} recover a suppression in the vertical component of the viscosity that scales most closely to linear.

Taking radial profiles for the mixing length and convective velocity from our fiducial models of WASP-12, 
assuming Zahn's scaling of the viscosity, and integrating over the convective envelope,  yields $t_{\rm diss} \approx 300$ yr.
From the rate of dissipation, an estimate for the orbital semi-major axis $a\approx 0.0234\unit{au}$ and the stellar moment of inertia $I_\ast$, 
one can also estimate the synchronization time \citep{Zahn2013} 
\begin{equation}
\frac{1}{t_{\rm sync}} = \frac{1}{t_{\rm diss}}\frac{m_{\mathrm{b}}^2 R_\ast^2}{M_\ast I_\ast}\left(\frac{R_\ast}{a}\right)^6,
\end{equation}
which yields $t_{\rm sync} \approx 11\unit{Gyr}$.
This suggests that viscous dissipation of the equilibrium tide is too weak to have significantly spun up the star, 
a result consistent with observations assuming a low initial rotation rate.  The corresponding orbital decay rate however,
\begin{equation}
\frac{P}{\dot{P}} = \left(\frac{m_{\mathrm{b}} a^2}{2I_\ast}\right)t_{\rm sync}\approx 1.2 \unit{Gyr},
\end{equation}
is several orders of magnitude too long to explain the observed decay.

\section{Dynamical Tide}\label{sec:dyn}
In addition to the hydrostatic tidal response of the equilibrium tide, there must 
also exist a low frequency dynamical response that mathematically arises from a condition for the continuity of fluid displacements across the 
radiative-convective boundary of the star.  Dubbed the dynamical tide, this fluid response results in the excitation of internal gravity waves at the star's 
radiative-convective boundary that propagate inwards to be damped by radiative diffusion.  The dynamical tide couples to the star's natural 
eigenfrequencies, potentially dissipating the tide at a rate orders of magnitude above the equilibrium rate if the system lies close to resonance.  
Provided the damping mechanisms acting on the dynamical tide are efficient to the point where waves are being damped before returning to the 
radiative-convective boundary, the resonances are broadened to the point of overlap.  Under the assumption that the resonances overlap, the 
dissipation rate is estimated as a frequency average over the resonances.  Adapting an expression for the frequency-averaged torque $\bar{\tau}$
from \citet{2017MNRAS.467.2146K}, to WASP-12's outer convective boundary and using quantities obtained from our fiducial models,
\begin{equation}
\begin{split}\label{eq:kushnir}
\bar{\tau}&\approx \frac{2Gm_{\mathrm{b}}^2}{R_{+}}\left(\frac{R_{+}}{a}\right)^6\left(\frac{R_{+}^3}{GM_{+}}\right)^{1/2}\frac{\rho_{+}}{\bar{\rho}_{+}}
\left(1 - \frac{\rho_{+}}{\bar{\rho}_{+}}\right)^2\omega_{{\rm orb}}\\
&\approx 2\times 10^{-7}\left(\frac{Gm_{\mathrm{b}}^2}{a}\right)
\end{split}
\end{equation}
where $\rho_{+}$ is the mass density at $R_{+}$, $\bar{\rho}_{+}$ is the mean mass density interior to $R_{+}$, $M_{+}$ is the mass interior 
to $R_{+}$, and $\omega_{\rm orb}= 6.67\times 10^{-5}\unit{s^{-1}}$ is the orbital 
angular frequency.  
Though there is an analogous torque caused by waves excited at the inner 
radiative-convective boundary for model A, the frequency averaged torque is some seven orders of magnitude smaller as $\bar{\tau} \propto r^{13/2}$.  
The circular orbital decay rate corresponding to the above torque is,
\begin{equation}
\frac{\dot{a}}{a} = -\frac{2\bar{\tau}}{m_{\mathrm{b}}\sqrt{GM_\ast a}} \approx -\frac{1}{1 \unit{Myr}}
\end{equation} 
which is fairly close to the observationally inferred decay rate $\dot{a}/a=\left(4.8 \unit{Myr}\right)^{-1}$.  This picture of tidal decay via the dynamical tide 
is a natural explanation for any models with a radiative core such as B and C.  Tidally excited internal gravity waves would freely propagate inwards and
 break near the center of the star, resulting in the above decay rate.  But as these models have their drawbacks (see \S\ref{sec:models}),
 we now ask whether models with a convective core such as model A have a mechanism to recover the frequency-averaged torque.

\subsection{Damping rates}\label{sec:damp}
The frequency-averaged torque can be recovered by model A if the radiative diffusion timescale or the viscous damping timescale in the convective
envelope is comparable to the propagation time for a gravity wave.
Because the Kelvin-Helmholtz timescale in roughly 
solar mass stars is relatively long, we operate under a quasi-adiabatic assumption in calculating these damping rates.
The relevant linearized work integral for calculating the radiative diffusion timescale is,
\begin{equation}
W \approx -\int_{R_{-}}^{R_{+}} \frac{\delta T}{T}\left( \pmb{\nabla} \cdot \pmb{F}_1\right) d^3 \pmb{r},
\end{equation}
where $\delta T$ is the Lagrangian perturbation to temperature and the Lagrangian heat flux perturbation $\delta \pmb{F}$ is replaced with $\pmb{F}_1$, 
its Eulerian perturbation, because the star is approximately in nuclear equilibrium on timescales short compared to the main-sequence lifetime.  
We allow the integral to range from the inner radiative-convective boundary ($R_-$) to the outer one ($R_+$) rather than the whole of the star.
Though there is a small positive contribution to the work integral from the convective regions, the contribution is small
because the work integral ends up being proportional to the superadiabatic gradient $\left(\nabla_{{\rm ad}} - \nabla\right)\ll1$.
Because the wavelength of these modes is small compared to a pressure scale height $H_p$, the opacity can be approximated as roughly 
constant and the above integral simplifies to,
\begin{equation}
W \approx  \int_{R_{-}}^{R_{+}}KT \nabla_{{\rm ad}}\left(\nabla_{{\rm ad}} - \nabla\right)\left[\left|\frac{d\xi_r}{dr}\right|^2 + \frac{l(l+1)}{r^2}\left|\xi_r\right|^2\right]dr,
\end{equation}
where $l$ is a mode's angular order, $K$ is the thermal conductivity, $\nabla$ is the temperature gradient, 
and $\nabla_{{\rm ad}}$ is the adiabatic temperature gradient.
To determine the linear eigenfunctions $\xi_r,\xi_h$ we numerically integrated the well known fourth-order set of stellar structure equations for linear, adiabatic,
non-radial perturbations by shooting to a fitting point at the outer radiative-convective boundary.
This yielded a radiative damping rate
$\gamma_{\rm rad} \approx \left(300\text{ yr}\right)^{-1}$.  Because this is orders of magnitude slower than the propagation time
\begin{equation}
t_{\rm prop}=\int \left|\frac{\partial k_r}{\partial \omega}\right|dr\approx \frac{\sqrt{l(l+1)}}{\omega^2} \int_{R_-}^{R_+}\frac{N}{r} dr\approx 9\unit{d},
\end{equation}
radiative diffusion by itself is not significant enough to broaden the resonant peaks to the point of overlap.

Although convective viscosity is not effective at dissipating the equilibrium tide in this system, convective viscosity could also damp 
internal gravity waves as they evanesce in convective regions.  
The total viscous diffusion associated with shear tensor $\sigma_{ij}$ and dynamic viscosity $\mu$ in Einstein notation is 
\begin{equation}
\dot{E}_{\rm visc} = \int\mu\left(\sigma_{ij}^2  -\frac{1}{3} \sigma_{ii}^2\right) d^3 \pmb{r}\approx\int\mu \sigma_{ij}^2 d^3 \pmb{r}.
\end{equation}
Solving for the squared components of the shear tensor in spherical polar coordinates, we find that the viscous work due to convection is of the form,
\begin{multline}
\dot{E}_{\rm visc} = \omega^2\int r^2 dr \mu\left[\left|\frac{d\xi_r}{dr}\right|^2 + 2\left(l^2 +l + 1\right)r^{-2}\left|\xi_r\right|^2\right.\\
\left.- 5l(l+1)r^{-2}\Re\left(\xi_h^\ast\xi_r\right)+l(l+1)(l^2+l+1)r^{-2}\left|\xi_h\right| ^2\vphantom{\left|\frac{d\xi_r}{dr}\right|^2}\right].
\end{multline}
For model A, this work integral corresponds to a damping rate of $\gamma_{\rm visc} \approx \left(300\text{ yr}\right)^{-1}$, still substantially long to inhibit averaging over the resonances.  

Though we can't justify the use of a frequency averaged dissipation to explain the observed decay, 
it's possible that we are observing this system sufficiently close to resonance to produce a high decay rate.  Adopting a damping rate,
$1/\gamma \equiv 1/\gamma_{\rm rad} + 1/\gamma_{\rm visc}$,
a circular orbit (consistent with observations) and uniform observation in time, the probability of seeing the system 
with observed decay rate with its $1\sigma$ errors, $-32 <\dot{P}<-26 \unit{ms\,yr^{-1}}$, is at the level of $\approx 10^{-7}$. Of course this probability should not 
be accepted in a rigorous sense because it ignores selection biases, 
but it is still instructive to share how truly little the resonances are broadened by our selected damping mechanisms.

\begin{figure}
\centering
\includegraphics[width=\linewidth]{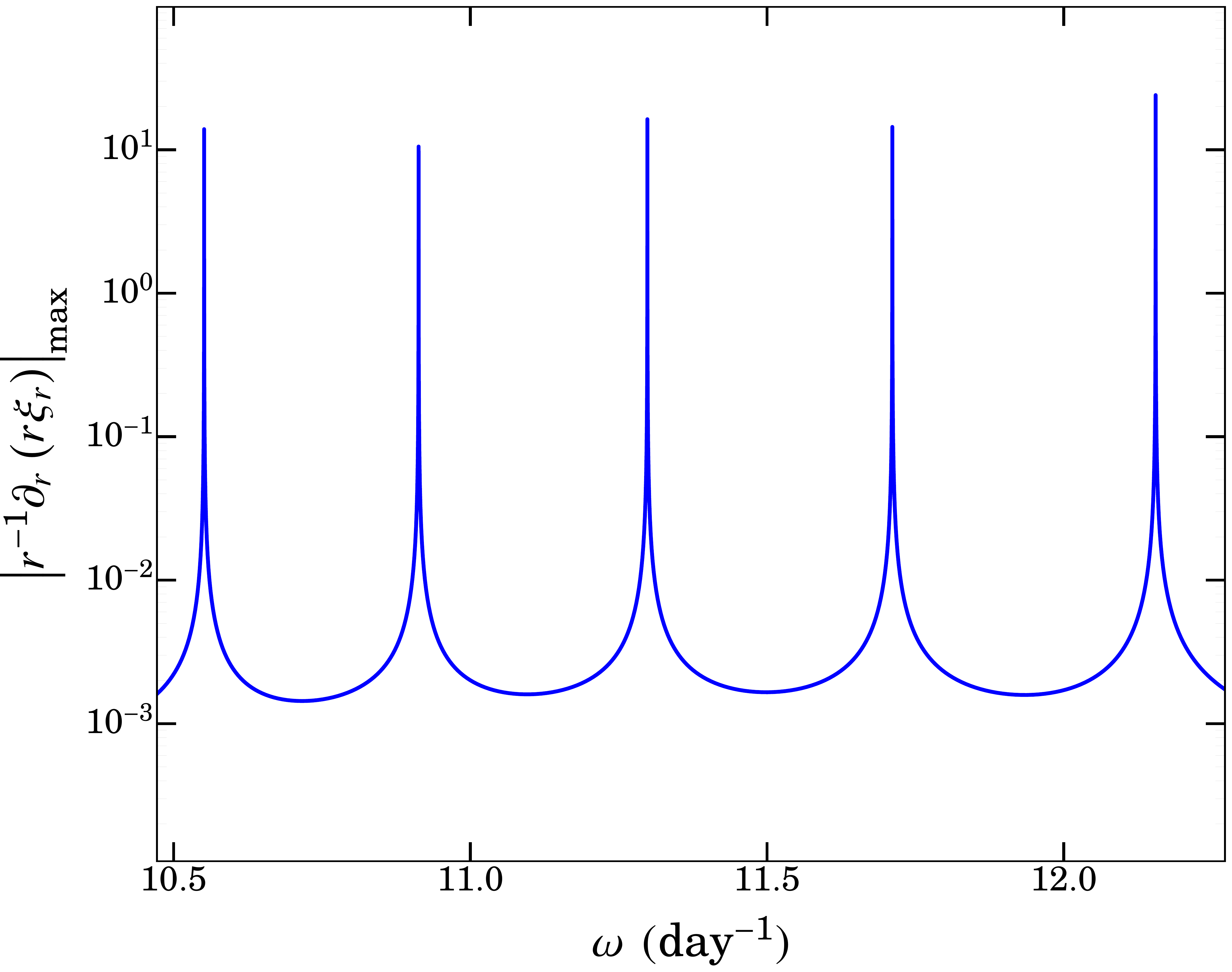}
\caption{The breaking criterion in linear theory as a function of tidal frequency.  The wave can sufficiently overturn stratification and break when $\left|r^{-1}\partial_r\left(r\xi_r\right)\right|_{\rm max}>1$}
\label{fig:strain}
\end{figure}

Without an effective damping mechanism, it is still possible to recover a frequency-averaged decay rate if the modes excited in the star 
are able to overturn stratification and break at some radius in their zone of propagation, thus depositing all their energy.  In linear theory, 
this criterion for breaking is simply $\Delta\equiv r^{-1}\partial_r\left(r\xi_r\right) >1$.  
Fig. \ref{fig:strain} shows the maximal value of $\Delta$ in model A for a range of frequencies.  
Because $\Delta_{\rm max} >1$ only for $\omega$ close to resonance, wave-breaking in model A does not provide a natural explanation for the tidal decay.

\subsection{Rotational effects}\label{sec:spinup}
\begin{figure}
\centering
\includegraphics[width=\linewidth]{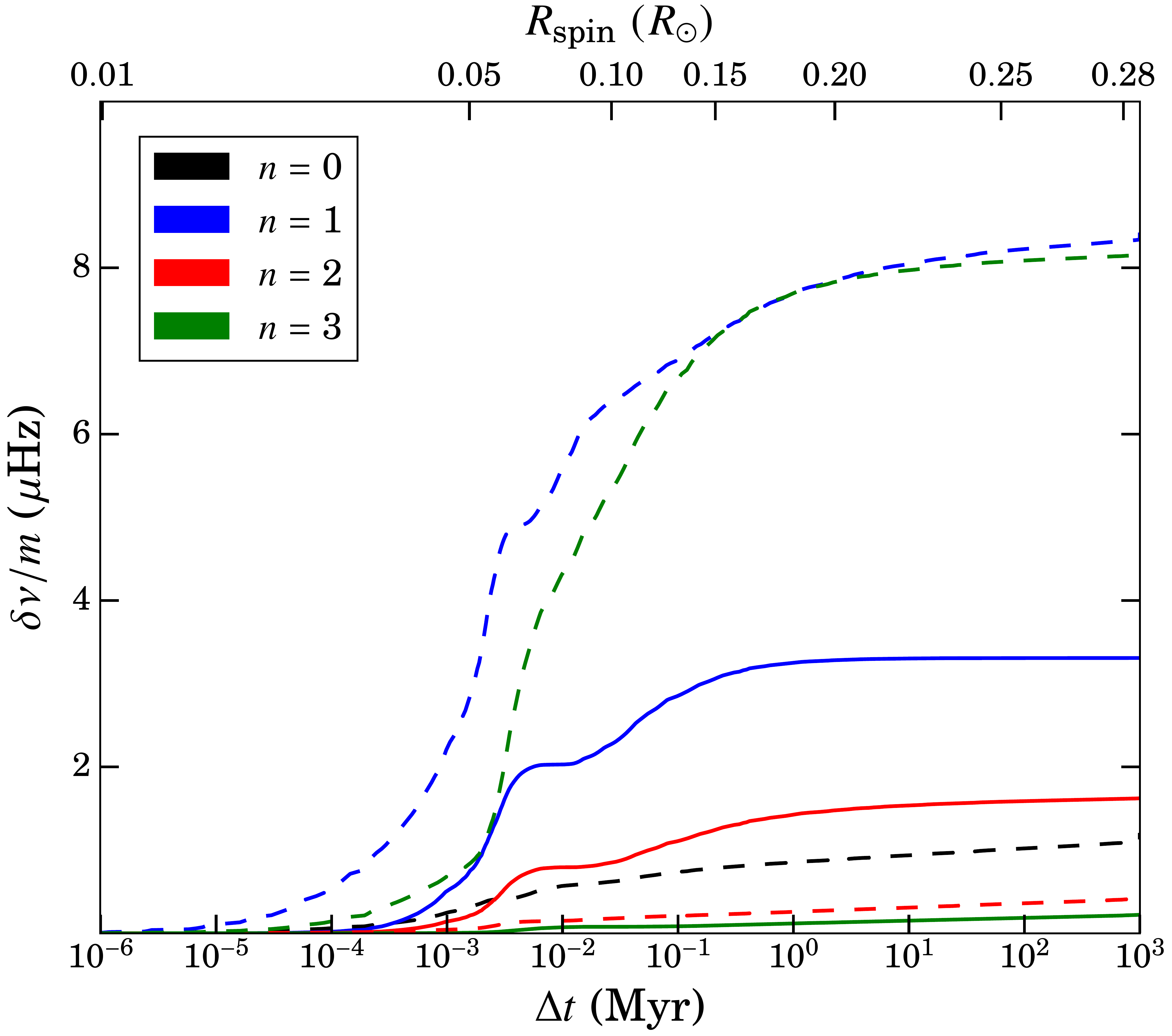}
\caption{Low-order frequency splittings due to a synchronously spinning stellar core  versus its radius
  $R_{\rm spin}$, or equivalently, versus the duration $\Delta t$ of the tidal torque.  Color denotes the radial order for a
  mode while line style denotes the angular order $l=1$ (solid) or $l=2$ (dashed).}
\label{fig:spinup}
\end{figure}

Assuming that WASP-12b's signature is indeed due to decay via the dynamical tide, as is suspected for model B or C, the extent 
to which the star should have been spun up to synchronous rotation can be explicitly calculated.  Even though WASP-12 is observed to have small 
surface rotation, at least some part of the core of the star should have been spun up from internal gravity waves breaking 
and depositing their angular momentum.  We make the approximation that the relevant torque is changed appreciably only by changes
to the orbit and not by changes to the star itself, so that $\tau \propto a^\eta$.  Given the period of time which the dynamical tide has been acting $\Delta t$,
the moment of inertia of the synchronously rotating core is,
\begin{equation}
I_{\rm spin} = m_{\mathrm{b}} a^2\left[\left(1+\frac{\Delta t (1-2\eta)\tau}{m_{\mathrm{b}}\sqrt{GM_\ast a}}\right)^\frac{1}{1-2\eta}-1\right],
\end{equation}
where the $\tau$ and $a$ refer to present day values.  In the following,
we adopt the form of the torque in eq. \ref{eq:kushnir} ($\eta=-10$) and use model B to estimate relevant stellar quantities.
For model B, the time between the best fit model and the disappearance of a convective core is $6\unit{Myr}\gtrsim\Delta t \gtrsim 3 \unit{Myr}$.  
But because the precursor to model B that has not yet lost its convective core still manages to fit the observables well, the actual $\Delta t$ could be 
made arbitrarily small and it's better to take $6\unit{Myr}\gtrsim\Delta t \gtrsim 0 \unit{Myr}$ .  
On the one hand, we could be seeing this system 100 years after the convective core disappeared, in which case the the core has not been spun up 
significantly, but 
probabilistically it's most likely we're seeing this system on the order of millions of years after the convective core disappeared.  
Even models 10-100 Myr after 
the core disappeared don't fit the observables terribly, but using one of these values doesn't change the radial extent of the spun up core due to the steep 
dependence of $\tau$ on $a$.  This fact is shown in Fig. \ref{fig:spinup} where we scale the upper abcissa with the value of
 the radius of the synchronously rotating core $R_{\rm spin}$ to the corresponding $\Delta t$ on the lower abcissa.  This insensitivity of $R_{\rm spin}$ on $\Delta t$ provides a potentially testable prediction of the dynamical tide explanation--if WASP-12b's decay
is an effect of the dynamical tide, the innnermost $\approx 0.2 R_\odot$ of WASP-12 itself should be rapidly rotating.  

Because the linear rotational frequency of this core $\nu_{\rm rot} = 66\unit{\mu Hz}$ is significantly less than the linear eigenfrequencies of 
our subgiant model, the rotational splittings can be estimated in a perturbative manner as in \citet{2010aste.book.....A}.  
For azimuthal order $m$, the splittings $\delta\nu$ can be written: 
\begin{equation}
\delta\nu = m\int _0^{R_\ast} K_{nl}(r)\nu_{\rm rot}(r) dr,
\end{equation}
where $K_{nl}$ is the unnormalized rotational kernel for radial order $n$, angular order $l$,
\begin{equation}
K_{nl} \equiv \frac{\int_0^{R_\ast}\left[\xi_r^2+l(l+1)\xi_h^2-2\xi_r\xi_h-\xi_h^2\right]\rho r^2 dr}{\int_0^{R_\ast}\left[\xi_r^2+l(l+1)\xi_h^2\right]\rho r^2 dr}.
\end{equation}
We present several of these low-order rotational splittings in Fig. \ref{fig:spinup} as a function of the size of the spinning core, and find that they are on the order
of a few $\mu$Hz.  Compare this to the corresponding rotational splittings in a model rotating uniformly at the measured $v\sin i_\ast = 1.6$ km s$^{-1}$; though the splittings would be enhanced by non-zero contributions from the entire star, ultimately they would remain orders of magnitude smaller owing to 
a much lower $\nu_{\rm rot}$.

\section{Discussion}\label{sec:disc}

We have seen that the apparent period change ($\dot P$) observed in the transits of WASP-12b cannot easily be explained as secular
orbital decay.  Standard mechanisms of tidal dissipation are too slow, unless the orbit happens to be close to resonance
with a global g~mode.  We have estimated the probability for this to be quite small.  

The leading alternative explanation for the anomalous transit times is that the planetary orbit is slightly eccentric,
$e\approx 2\times 10^{-3}$.  In this interpretation, the true orbital period is constant, but the transit times depart
slightly from a linear ephemeris due to precession of periastron at a rate $\dot\omega\approx 26\unit{deg\,yr^{-1}}$
\citep{Maciejewski+2016,Patra+2017}.  The latter authors estimate that for a reasonable tidal quality factor of
the planet itself, $Q_{\rm p}\le 10^6$, any primordial eccentricity would have decayed to $e < 10^{-3}$ after a few
million years, whereas the system age appears to be $> 1\unit{Gyr}$.  Therefore, the eccentricity would have
to be recently excited or continually forced.  

We now briefly examine mechanisms for forcing the eccentricity or modulating the period of the orbit via changes
in host star or third bodies.

In the following, unless otherwise noted, we take $M_*=1.4\,M_\odot$, which
is slightly higher than any of the values in Table~\ref{tab:models}.  Therefore $R_*=1.64\,R_\odot$ based on the mean
density adopted in Table~\ref{tab:properties}.  With \cite{Southworth2012}'s result that
$R_{\rm b}/R_*\approx0.1159\pm0.0033$, we then have $R_{\rm b}\approx 1.89 R_{\rm J}$ for the planetary radius.  Both radii
would scale $\propto M_*^{1/3}$ to other assumed values of the stellar mass.  Adopting the radial velocity amplitude
$K= 221.9\pm3.1\unit{m\,s^{-1}}$ from \cite{Knutson+2014} and the inclination $I=(83\pm0.5)^\circ$ from
\cite{Maciejewski+2013}, and since $m_{\mathrm{b}}(M_*+m_{\mathrm{b}})^{-2/3}= K_*\sec I (P/2\pi G)^{1/3}$, we then have
$m_{\mathrm{b}}=1.41\,M_{\rm J}$; this scales approximately as $M_*^{1/3}$.  Finally the semimajor axis becomes
$a=(P/2\pi)^{2/3}[G(M_*+m_{\mathrm{b}})]^{1/3}\approx 0.02322\unit{au}$.

\subsection{Eccentricity from convection}

\citet{Phinney1992} proposed that the small measured eccentricities of binary millisecond pulsars with white-dwarf
companions can be explained by potential fluctuations associated with convection in the envelope of the companion
when on the giant or asymptotic-giant branch.  With few exceptions, the orbits of subsequently discovered binary millisecond
pulsars have conformed well to the predictions of this model \citep{Lorimer2008}.

Adapted to the WASP-12 system, so that the reduced mass $\mu\approx 1.4\,M_{\rm J}$, Phinney's equation (7.33) reads
\begin{equation}
  \label{eq:ePhinney}
  \langle e^2\rangle^{1/2} \approx 2\times 10^{-5}  \left(\frac{L_*R_{\rm env}}{5\,L_\odot R_\odot}\cdot
\frac{1.4 M_\odot}{M_*}\right)^{1/3} \left(\frac{M_{\rm env}}{0.0004\,M_\odot}\right)^{1/6}\,,
\end{equation}
in which $R_{\rm env}\approx 1.4\,R_\odot$ is the radius at the base of the outer convection zone in our preferred model
for WASP-12. $M_{\rm env}$, the mass of that zone, is sensitive to the effective temperature, metallicity, and
evolutionary state of the star, but in view of the sixth root, no plausible value of $M_{\rm env}$ could make up the two
orders of magnitude by which the r.m.s. eccentricity predicted by eq.~\eqref{eq:ePhinney} falls short of the value
required to explain the quadratic term in the transit ephemeris.  Furthermore, as Phinney remarks, his eq.~(7.33)
probably overestimates the eccentricity expected when the turnover time of the largest convective eddies exceeds the
tidal period, as occurs in WASP-12 by at least one order of magnitude.

\subsection{The Applegate effect}

\citet{Applegate+Patterson1987} and \citet{Applegate1992} suggested that long-term modulations observed in the eclipse
times of some close stellar binaries, including V471~Tau and Algol, are caused by slow changes in the quadrupole moment
of one or both stars induced by their magnetic cycles.  In the later version of this idea, the magnetic stress is not
large enough to distort the equilibrium shape of the star directly, but rather slowly redistributes angular momentum
within the star(s), leading to changes in the rotationally-induced quadrupole.  Because the changes are slow, they would
not excite the eccentricity of the orbit, but the quadrupole contributes to the central force between the stars and
hence to the orbital period itself.  \citet[hereafter WM10]{Watson+Marsh2010} have scaled \citet{Applegate1992}'s model
to several exoplanet systems.  For WASP-12b, they estimate that the anomaly in the transit time ($O-C$, observed minus
calculated) could be as much as $42~(T/50\,\mathrm{yr})^{3/2}$, where $T$ is the period on which the dynamo modulates
the internal differential rotation.  This last could be the same as the period of the magnetic dipole, or half that,
depending on the type of dynamo.

MW10's predicted variation is not a great deal smaller than the $\sim$2-minute departure from a linear transit
ephemeris found by \citet{Patra+2017}.  It depends on several several uncertain parameters besides the dynamo period
$T$, so one ought to consider whether the uncertainties in these parameters might allow the Applegate effect to explain
the WASP 12 data.  The relevant parameters are the rotation period of the star, for which WM10 take
$P_{\rm rot}=36\,\mathrm{d}$, the fractional mass of the convection zone, for which they take $M_{\rm env}/M_*=0.1$, and
the portion of the mean luminosity that is converted to mechanical form to change the differential rotation.  For the
latter they take $\Delta L=0.1\,L$; this seems large, but perhaps not in direct conflict with observations because, as
they point out, the luminosity variation at the photosphere could be much smaller due to the
thermal inertia of the convection zone (i.e., the ratio of its total thermal energy to the luminosity of the star; this
is about 300~yr for WASP 12).  WM10's equations imply that the transit-time anomalies scale with these parameters as
follows:
\begin{equation}
  \label{eq:MWscalings}
(O-C)_{\rm max}\propto T^{3/2}P_{\rm rot}^{-1}\left(\frac{M_{\rm env}}{M_*}\right)^{1/2}(\Delta L)^{1/2}\,.
\end{equation}
The mass of the convective envelope of WASP~12 is probably $\lesssim 10^{-3}\,M_*$, as remarked above; following
eq~\eqref{eq:MWscalings}, this would reduce the predicted $O-C$ by an order of magnitude.  On the other hand, the
rotation period may be rather less than the assumed value if the star is viewed near pole on, as Rossiter-MacLaughlin
measurements suggest \citep{2012ApJ...757...18A}.  The median rotation period for main-sequence F8
stars\footnote{\citet{Hebb+2009} classify WASP 12 as F9V} is $\approx 8\,\mathrm{d}$ \citep{Nielsen+2013}.  Since dynamo
periods appear to correlate positively with stellar rotation periods \citep{Saar+Brandenburg1999,Boehm-Vitense2007},
however, the positive scaling with $T$ seems likely to overwhelm the negative scaling with $P_{\rm rot}$ in
eq.~\eqref{eq:MWscalings}.

If MW10's scalings are applied to the Sun,
they predict a variation $\Delta J_2\gtrsim 5\times 10^{-8}$ in its rotationally-induced
dimensionless quadrupole moment over the dynamo cycle.  The internal differential rotation of the Sun has been
directly constrained by helioseismology, and for a significant fraction of a cycle.  \citet{Antia+2008} have used these
data to estimate that $\langle J_2\rangle_\odot=2.2\pm0.01\times 10^{-7}$, and the variation over a nine-year period
to be $\lesssim 1\times 10^{-10}$, i.e. several orders of magnitude smaller than MW10's assumptions would predict.

For these reasons (i.e., both our estimates of the actual parameters of WASP-12, as well as comparision with
heliosesimological inferences for the Sun), it is unlikely that the Applegate effect explains the transit-time anomolies
of WASP-12b.

\subsection{Bow shock}

Ultraviolet absorption is seen just before each transit of WASP-12b and has been
interpreted as evidence for mass loss from the planet through its inner Lagrange point \citep{Fossati+2010a}.  Alternatively,
this could be the signature of a bow shock ahead of the planet encountering a wind from the star
\citep{Lai+2010,Vidotto+2010}.   Such a shock would exert a drag on WASP-12b's
orbit. As shown here, however, an improbably dense wind would be required to explain the observed $\dot P$.

The torque exerted on the planet by the shock is 
$C_D\pi R_{\mathrm{b}}^2\rho_{\mathrm{w}} (v_{\mathrm{b}}^2+v_{\mathrm{w}}^2)^{1/2}v_{\mathrm{b}}a$, 
in which $C_D$ is a factor of order unity (the drag coefficient), $\rho_{\mathrm{w}}$ the
pre-shock density of the wind, $v_{\mathrm{w}}$ the wind velocity, $R_{\mathrm{b}}\approx 1.9\,\mathrm{R_{\rm J}}$ the radius of
the planet, and $v_{\mathrm{b}}\approx (GM_*/a)^{1/2}$ the orbital velocity.  The decay
timescale is then
\begin{equation}
  \label{eq:drag}
  \frac{P}{\dot P} = \frac{m_{\mathrm{b}}}{3\pi C_D R_{\mathrm{b}}^2\rho_{\mathrm{w}}(v_{\mathrm{b}}^2+v_{\mathrm{w}}^2)^{1/2}}
 \approx 4\times10^{12}\unit{yr}.
\end{equation}
For the numerical estimate, we have taken $C_D=0.3$, and $\rho_{\mathrm{w}}=2\times10^{-18}\unit{g\,cm^{-3}}$
(i.e. $n_{\mathrm{H}}=1.5\times10^6\unit{cm^{-3}}$); the latter follows \cite{Vidotto+2010} and
implies a stellar mass-loss rate of $10^{-12.3}(v_{\rm wind}/100\unit{km\,s^{-1}})\unit{M_\odot\,yr^{-1}}$.  In order to
explain the apparent decay rate ($P/\dot P\approx 3\unit{Myr}$), the wind density would have to increase some six
orders of magnitude, making the mass-loss timescale of the star $\lesssim 10\unit{Myr}$.  This is unreasonable as the
star is probably older than $1\unit{Gyr}$.

\subsection{Kozai-Lidov oscillations}\label{subsec:KL}

We consider the possibility that a non-transiting third body in the system continuously excites a small eccentricity
in the orbit of WASP-12b so that, as suggested by \citet{Maciejewski+2016}, the transit-time anomolies result from
apsidal precession of the slightly elliptical orbit.

Apsidal precession itself imposes a lower bound on the perturbations that such a hypothetical companion must exert to
excite WASP-12b's eccentricity.  Let the companion have mass $m_{\mathrm{c}}$, semimajor axis $a_{\mathrm{c}}$, and
orbital eccentricity $e_{\mathrm{c}}$, and let $\{m_{\mathrm{b}},a_{\mathrm{b}},e_{\mathrm{b}}\}$ be those of WASP-12b
itself.  By a standard calculation in secular perturbation theory, one can show that if $e_{\mathrm{b}}\ll 1$ initially,
then $e_{\mathrm{b}}$ will grow by the Kozai-Lidov mechanism (hereafter KLM) only if
\begin{equation}
  \label{eq:KLlimit}
  \frac{m_{\mathrm{c}}}{a_{\mathrm{c}}^3(1-e_{\mathrm{c}}^2)^{3/2}} > \frac{10}{3} k_{2\mathrm{b}}
\frac{M_*^2R_{\mathrm{b}}^5}{m_{\mathrm{b}} a_{\mathrm{b}}^8}\,,
\end{equation}
in which $R_{\mathrm{b}}$ is the radius of WASP-12b and $k_{2\mathrm{b}}$ its Love number, these two quantities being important
for the apsidal precession rate.  The inequality \eqref{eq:KLlimit} assumes that the
orbital planes of $m_{\mathrm{c}}$ and $m_{\mathrm{b}}$ are orthogonal, which maximizes the efficiency of the KLM.  We are also
assuming $a_{\mathrm{c}}>a_{\mathrm{b}}$, i.e. the third body's orbit is exterior to that of WASP-12b.  The orbits
should not cross, whence $a_{\mathrm{c}}(1-e_{\mathrm{c}})>a_{\mathrm{b}}$, and therefore
$a_{\mathrm{c}}(1-e_{\mathrm{c}}^2)^{1/2}>\sqrt{a_{\mathrm{b}}a_{\mathrm{c}}}$.
With $k_{2\mathrm{b}}\approx0.6$, the
lower bound on the companion's mass for the KLM becomes
\begin{equation}
  \label{eq:KL2}
  m_{\mathrm{c}} > 77. \left(\frac{a_{\mathrm{c}}}{a_{\mathrm{b}}}\right)^{3/2}\unit{M_\oplus}
\end{equation}

An upper bound on $m_{\mathrm{c}}$ follows from the published
radial-velocity data \citep{Hebb+2009,Husnoo+2011,2012ApJ...757...18A,Bonomo+2017}.  After subtraction of the WASP~12b 
signal\footnote{We subtract an optimally scaled multiple of the photometric ephemeris of \cite{Patra+2017}, including
their secular period derivative $\dot P= (0.92\pm0.01)\times 10^{-9}$.  Thus this limit applies to companions with
periods less than the span, of the data, $\sim 7$~yr.}
and correction for the nominal measurement errors, these data have variance $\approx (9\unit{m\, s^{-1}})^2$.  The
RV signal of the hypothetical WASP-12c should be no larger than this.
Therefore
\begin{equation}
  \label{eq:mcrv}
  m_{\mathrm{c}} < 18\,f^{-1/2} \left(\frac{a_{\mathrm{c}}}{a_{\mathrm{b}}}\right)^{1/2}\unit{M_\oplus}\,,
\end{equation}
with $f$ being a geometrical factor that determines the mean-square projection of the orbital velocity onto the line of
sight:
\begin{equation}
  \label{eq:ffac}
  f(e_{\mathrm{c}},\omega_{\mathrm{c}},I_{\mathrm{c}}) = \frac{\sin^2\omega_{\mathrm{c}}
+\sqrt{1-e^2_{\mathrm{c}}}\,\cos^2\omega_{\mathrm{c}}}{1+\sqrt{1-e^2_{\mathrm{c}}}}\sin^2I_{\mathrm{c}}\,.
\end{equation}

In order that the KLM operate, the relative inclination of the two planetary orbits must be greater than 
$\sin^{-1}\sqrt{2/5}\approx 39.2^\circ$, so
\begin{equation*}
  \cos I_{\mathrm{c}}\cos I_{\mathrm{b}}  +\sin I_{\mathrm{c}} \sin I_{\mathrm{b}}\cos(\Omega_{\mathrm{c}}-\Omega_{\mathrm{b}})
< \sqrt{3/5}
\end{equation*}
with $\Omega_{\mathrm{b,c}}$ being the longitudes of the ascending nodes.  Since the inclination of WASP~12b is
$I_{\mathrm{b}}\approx (83\pm0.5)^\circ$ \citep{Maciejewski+2013}, the above constraint is compatible with
$I_{\mathrm{c}}\approx 0$, and of course also with any eccentricity $e_{\mathrm{c}}$ or argument of periastron
$\omega_{\mathrm{c}}$.  So the factor $f$ could be arbitrarily small.  The two inequalities \eqref{eq:KL2} \&
\eqref{eq:mcrv} could therefore both be satisfied by an exterior perturber ($a_{\mathrm{c}}>a_{\mathrm{b}}$), although
this becomes less probable as the separation between the orbits increases because of the different scalings with
$a_{\mathrm{c}}/a_{\mathrm{b}}$.  Furthermore, eqs.~\eqref{eq:mcrv}-\eqref{eq:ffac} suppose that the radial velocity is
measured continuously, whereas in fact it is sampled somewhat sparsely and irregularly: nearly half of the $\sim 90$
measurements were made by \citet{2012ApJ...757...18A} in a single night.  If WASP~12c's orbit were
highly eccentric, and thus hovering usually near apastron, its full radial-velocity amplitude might not be sampled.

We have not systematically investigated the probability that both of the mutually antagonistic bounds 
\eqref{eq:KL2} and \eqref{eq:mcrv} could be satisfied.  Nevertheless, the Kozai-Lidov mechanism does not
seem to provide a natural explanation for the quasi-secular transit-time anomalies of WASP~12b.
The hypothesis is attractive only in comparison to all of the other possibilities that we have investigated.

\subsection{Resonance}

We have considered the possibility that the orbital variations of WASP~12b are caused by resonant interactions with
an unseen planet.  We focus on mean-motion resonances. 

Suppose first a 1:1 resonance, in other words, a small trojan planet librating around the stable Lagrange points of
the WASP~12+WASP~12b system.\footnote{We thank Scott Tremaine for suggesting that we look into this.}
  The inferred amplitude of the period variation is $29\pm3\unit{ms\,yr^{-1}}$
\citep{Patra+2017}, amounting to $\Delta\ln P\approx 3\times 10^{-6}$ over the 9 years that transits have been
monitored.  We estimate that a roughly lunar mass in a ``horseshoe''  1:1 resonant libration
could modulate WASP~12b's period at this amplitude.
This would easily satisfy the limit $m_{\mathrm c}<34\,\mathrm{M}_\odot$ on Trojan companions to WASP-12b
found by \cite{Lillo-Box+2018}, who based their analysis on archival radial velocities.
The difficulty, however, is in the period of the modulation.
It is well known that small-amplitude librations around the Lagrange points in the coplanar restricted three-body
problem have period $P_{\rm lib} = P_{\rm orb}\times 2(1+q)/\sqrt{27 q}$, where $P_{\rm orb}$ is the orbital period
of the massive bodies and $q<0.04$ is their mass ratio.  In the present case where $P_{\rm orb}=1.09\unit{d}$ and
$q\approx 10^{-3}$, $P_{\rm lib}\approx 13\unit{d}$.  A large-amplitude libration can have a somewhat longer
period than this, but not by more than a factor $\sim 2$
unless very close to the separatrix between libration and circulation, as we have convinced ourselves by numerical
experiments.  Such a $P_{\rm lib}$ is far too short to mistaken for a secular trend over 9~yr unless severely aliased,
which seems unlikely in view of the density of transit observations [see the tabulation in \citet{Patra+2017}].

We have also examined first-order mean motion resonances $P_{\mathrm c}:P_{\mathrm b} \approx (j+1):j$, with $j\ge 1$
an integer.  Our analysis is restricted to coplanar, near-circular cases, but the main conclusions would probably be similar
even for strongly misaligned orbits.  The unseen body WASP-12c is presumed to be much less massive than WASP-12b.

Close to such a resonance, the $j^{\mathrm{th}}$ azimuthal harmonic of the potential of the orbit of b directly forces
the eccentricity of c's orbit ($e_{\mathrm{c}}$), and the $(j+1)^{\mathrm{th}}$ harmonic of c forces $e_{\mathrm{b}}$.
In the first case, or ``exterior'' resonance, $e_{\mathrm{b}}$ is neglected to leading order, while $e_{\mathrm{c}}$ is
neglected for the interior resonance \citep[e.g.][]{Murray+Dermott2000}.  The forced eccentricities depend not only on
the masses $m_{\mathrm{c}}$ and $m_{\mathrm{b}}$ but also on the distances from exact resonance; these differ because of
the unforced apsidal precession rates of the two planets.  As already noted in \S\ref{subsec:KL}, the apsidal precession
of b is dominated by its tidal distortion: $\varpi_{\mathrm{b}0}\approx 3.9\times10^{-4} n_{\mathrm b}$, with
$n_{\mathrm{b}}=2\pi P_{\mathrm{b}}^{-1}$ being its mean motion.  If c is a smaller body such as a super-earth, its
apsidal motion is dominated by the axisymmetric potential of b's orbit.  Near the 2:1 resonances, we estimate that
$\dot\varpi_{\mathrm{c}0}\approx 3.8\times10^{-4}n_{\mathrm{b}}$.  Because of the coincidence that
$\dot\varpi_{\mathrm{b}}\approx \dot\varpi_{\mathrm{b}}$, the slow frequencies that measure the distance from resonance,
namely $\nu_{\mathrm{b}}\equiv j n_{\mathrm{b}}-(j+1) n_{\mathrm{c}}-\dot\varpi_{\mathrm{b}0}$ and
$\nu_{\mathrm{c}}\equiv j n_{\mathrm{b}}-(j+1) n_{\mathrm{c}}-\dot\varpi_{\mathrm{c}0}$ will usually be nearly equal, at least
for the 2:1 resonances ($j=1$).

Tidal dissipation within the planets damps the forced eccentricity at the rate
\begin{equation}
  \label{eq:Qcirc}
  \gamma_{\rm p}\equiv 
-\left(\frac{d\ln e}{dt}\right)_{\rm tide}
 = \frac{63}{4Q'_{\mathrm{p}}}\frac{M_*}{m_{\mathrm{p}}}\left(\frac{R_{\mathrm{p}}}{a_{\mathrm{p}}}\right)^5 n_{\mathrm{p}}
\end{equation}
where $Q'_{\mathrm{p}}$ is the tidal quality factor of planet p corrected for its Love number.
On short
timescales $\sim \nu^{-1}$, an equilibrium holds between forcing and damping.  Secularly however,
at second order in eccentricity and first order in the damping rate \eqref{eq:Qcirc}, there is a transfer
of orbital energy and angular momentum between planets.  The transfer is always outward,
i.e. from b to c in our case, but in the proportion $\Delta E = n_{\mathrm{c}}\Delta J$ for the interior resonance (where
the orbit of c is approximated as circular), and $\Delta E = n_{\mathrm{b}}\Delta J$ for the exterior resonance (where
$\Delta e_{\mathrm{b}}$ is neglected).
The rate of transfer of angular momentum is related to the tidal dissipation rates $\mathcal{\dot E}_{\mathrm{b,c}}>0$ by
\begin{equation}
  \label{eq:dJdt}
  \frac{dJ_{\mathrm c}}{dt}=-  \frac{dJ_{\mathrm b}}{dt} =
  \frac{\mathcal{\dot E}_{\mathrm{b}}+\mathcal{\dot E}_{\mathrm{c}}}{n_{\mathrm{b}}-n_{\mathrm{c}}}
\end{equation}
The effect of this torque is to increase the slow frequencies
$\nu_{\mathrm{b}}$ and $\nu_{\mathrm{c}}$, and hence to increase the distance from resonance if these frequencies
are already positive.

If body c is a super-earth, we estimate that $\mathcal{\dot E}_{\mathrm{b}}>\mathcal{\dot E}_{\mathrm{c}}$ by a factor
of at least a few at the first few mean-motion resonances ($j\lesssim 6$):
\begin{equation}
  \label{eq:Edotb}
  \mathcal{\dot E}_{\mathrm{b}} = \frac{\gamma_{\mathrm{b}} m_{\mathrm{b}} A^2_{\mathrm{b}}/4}{\nu_{\mathrm{b}}^2+\gamma^2_{\mathrm{b}}}\,,
\end{equation}
where
\begin{equation}
  \label{eq:Ab}
  A_{\mathrm{b}} = \frac{Gm_{\mathrm{c}}}{a_{\mathrm{b}}a_{\mathrm{c}}}
\left[\frac{d}{d\ln\alpha} b_{1/2}^{(j+1)} (\alpha)+2j b_{1/2}^{(j+1)}(\alpha)\right]_{\alpha =a_{\mathrm{b}}/a_{\mathrm{c}}},
\end{equation}
in which the functions $b_{1/2}^{(j+1)}(\alpha)$ are the usual Laplace coefficients.  The last equation follows from
first-order epicyclic theory if the damping term is inserted by hand. 
(These equations also determine $\mathcal{\dot E}_{\mathrm{c}}$ if all subscripts ``b'' and ``c'' are interchanged and
$j+1$ is replaced by $j$.)

For definiteness, let us focus on the
2:1 resonance, $j=1$, so that $\nu_{\mathrm{b}}\approx\nu_{\mathrm{c}}$ by the numerical coincidence noted above.
Presuming that $m_{\mathrm{c}}\ll m_{\mathrm{b}}$, the increase in $\nu$ due to the torque \eqref{eq:dJdt} is dominated
by the change in the mean motion of c, but $dE_{\mathrm{c}}/dJ_{\mathrm{c}}\approx n_{\mathrm{c}}$ because dissipation
occurs mainly in body b.  Hence
$dn_{\mathrm{c}}/dJ_{\mathrm{c}}\approx -3/m_{\mathrm{c}}a^2_{\mathrm{c}}$.
In the relevant regime where $\nu\gg\gamma$, $d\nu/dt\propto\nu^{-2}$ because of the denominator in eq.~\eqref{eq:Ab},
the other terms in eqs.~\eqref{eq:Edotb}-\eqref{eq:Ab} being effectively constant when $|\nu|\ll n_{\mathrm{b,c}}$.
Integrating this relation with the constants included yields
\begin{equation}
  \label{eq:nuest}
 \nu\approx
  0.02\left(\frac{10^6}{Q'_{\mathrm{b}}}\frac{m_{\mathrm{c}}}{M_\oplus}\frac{T}{\mathrm{Gyr}}\right)^{1/3}\,n_{\mathrm{b}}\,,
\end{equation}
presuming that the system started from exact resonance at time $T$ in the past.

The quantities in parentheses in eq.~\eqref{eq:nuest} are uncertain, but because of the cube root, it is unlikely that
the distance from resonance ($\nu/n_{\mathrm{b}}$) is much less than $10^{-2}$.  Now at a $(j+1):j$ resonance,
the combination $(j+1)n_{\mathrm{c}}-jn_{\mathrm{b}}$ is the \emph{forced} apsidal precession rate,
$\dot\varpi_{\mathrm{b}}$.  Therefore $\nu_{\mathrm{b}}=\dot\varpi_{\mathrm{b0}}-\dot\varpi_{\mathrm{b}}$.
Since we have previously estimated that $\dot\varpi_{\mathrm{b0}}\approx 4\times 10^{-4}n_{\mathrm{b}}$, it follows from
eq.~\eqref{eq:nuest} that $\dot\varpi_{\mathrm{b}}<0$, with a period $\sim 50\times P_{\mathrm{b}}\approx
55\,\mathrm{d}$.  Thus while it is possible to choose $m_{\mathrm{c}}$ so that the amplitude of the forced eccentricity
$e_{\mathrm{b}}=2\times10^{-3}$, the period of the apsidal precession is much too rapid
to explain the observed quasi-secular $\dot P$.

\section{Summary}\label{sec:summary}

We have revisited the possible causes of WASP-12b's departure from a linear ephemeris.  Either the orbit is decaying, or
some dynamical perturbation maintains a small eccentricity and the apsides precess on some period longer than a decade.
We have considered various perturbations induced by unseen third bodies or distortions of the star WASP-12 itself,
but none is consistent with all of the observational constraints, at least not without fine tuning.  

The conclusion therefore seems inescapable that the orbit is indeed decaying, presumably because of tidal dissipation in
the star.  Indeed, the dynamical tide---computed for a circular orbit and a negligibly rotating star---naturally yields
an orbital lifetime comparable to what is inferred from transit timing.  But this requires that the star has evolved
onto the subgiant branch and lost its convective core, as \citet{Weinberg+2018} have suggested.  In that case, the
g~modes excited at the base of WASP-12's thin surface convection zone might be just strong enough to damp nonlinearly in
the core, which would broaden the g-mode resonances so that they overlapped.  If WASP-12 were still on the main-sequence
and still had its convective core, the resonances would be very sharp, and the orbit would have to be implausibly close
to resonance to explain the current rate of orbital evolution.  Alternatively, if the star had a \emph{rapidly rotating}
core, with a rotation period as short or shorter than the period of the orbit, then the tidally excited g-modes would be
absorbed at the critical (corotation) layer \citep{Barker+Ogilvie2010}; the torque applied by absorption of the ingoing
waves would then maintain the rapid rotation of the layer and presumably of the core beneath it.  This begs the question
how the core could have started out with such rapid rotation, however.  Moreover, unlike the subgiant hypothesis,
it does not naturally explain why the decay timescale is so much shorter than the age of the star.

The observational constraints on WASP-12 itself, when fit to theoretical models for its structure made with the
\textsc{mesa} code, favor a main-sequence star rather than a subgiant.  Actually, we have not been able to find
any \textsc{mesa} model that fits all of the observations comfortably: the spectroscopically inferred $T_{\rm eff}$ and [Fe/H]
are in tension with the luminosity inferred from the GAIA-DR2 distance and \citet{Stassun+2017}'s bolometric flux.
This problem would exist even if there were no evidence for orbital decay, though the transit light curves are essential
for constraining the star's mean density.

\bigskip
We thank Josh Winn for introducing us to this problem and for much helpful advice and conversation.

\section*{Acknowledgements}
This work has made use of data from the European Space Agency (ESA) mission
{\it Gaia} (\url{https://www.cosmos.esa.int/gaia}), processed by the {\it Gaia}
Data Processing and Analysis Consortium (DPAC,
\url{https://www.cosmos.esa.int/web/gaia/dpac/consortium}). Funding for the DPAC
has been provided by national institutions, in particular the institutions
participating in the {\it Gaia} Multilateral Agreement.


\bibliographystyle{mnras}
\bibliography{tide} 




\bsp	
\label{lastpage}
\end{document}